\documentclass[runningheads]{llncs}

\usepackage[utf8]{inputenc}
\usepackage[T1]{fontenc}
\usepackage{graphicx}
\usepackage{hyperref}

\usepackage{url}
\usepackage{booktabs}
\usepackage{amsfonts}
\usepackage{nicefrac}
\usepackage{microtype}
\usepackage[disable]{todonotes}
\usepackage{amsfonts}
\usepackage{amsmath}
\usepackage{amssymb}
\usepackage{subfig}\usepackage{forloop}
\newcounter{loopc}

\def\permille{\ensuremath{{}^\text{o}\mkern-5mu/\mkern-3mu_\text{oo}}}

\begin{document}
\title{ACE: A Novel Approach for the Statistical Analysis of Pairwise Connectivity}

\author{Georg Krempl\inst{1}\orcidID{0000-0002-4153-2594}\and
Daniel Kottke\inst{2}\orcidID{0000-0002-7870-6033}\and
Tuan Pham\inst{2}\orcidID{0000-0002-5102-5561}}
  
\authorrunning{G. Krempl et al.}
\institute{Utrecht University, 3584 CC Utrecht, The Netherlands 
\email{g.m.krempl@uu.nl}\\
\url{https://www.uu.nl/staff/GMKrempl}
\and
Kassel University, 34121 Kassel, Germany
\email{\{daniel.kottke,tuan.pham\}@uni-kassel.de}}

\maketitle

\begin{abstract}
Analysing correlations between streams of events is an important problem. It arises for example in Neurosciences, when the connectivity of neurons should be inferred from spike trains that record neurons' individual spiking activity. While recently some approaches for inferring delayed synaptic connections have been proposed, they are limited in the types of connectivities and delays they are able to handle, or require computation-intensive procedures.
This paper proposes a faster and more flexible approach for analysing such delayed correlated activity: a statistical approach for the \textbf{A}nalysis of \textbf{C}onnectivity in spiking \textbf{E}vents (ACE), based on the idea of hypothesis testing. It first computes for any pair of a source and a target neuron the inter-spike delays between subsequent source- and target-spikes.
Then, it derives a null model for the distribution of inter-spike delays for \emph{uncorrelated}~neurons. Finally, it compares the observed distribution of inter-spike delays to this null model and infers pairwise connectivity based on the Pearson's $\chi^2$ test statistic.
Thus, ACE is capable to detect connections with a priori unknown, non-discrete (and potentially large) inter-spike delays, which might vary between pairs of neurons. Since ACE works incrementally, it has potential for being used in online processing. 
In our experiments, we visualise the advantages of ACE in varying experimental scenarios (except for one special case) and in a state-of-the-art dataset which has been generated for neuro-scientific research under most realistic conditions.

\end{abstract}

\section{Introduction}\label{sec:intro}
An important problem in various applications is detecting correlations between streams of events. This is of particular importance in Neurosciences, where it arises for example when inferring the functional connectivity of neurons \cite{PerkelGersteinMoore1967B}. Given spike trains with recordings of the neuron's individual spike activity, the objective is to detect correlations between the spike activities of pairs or networks of neurons. Most of the 
existing approaches are designed for detection of correlated synchronous activity, considering solely events within the same discretised
time interval. More recently, the detection of delayed synaptic connections has gained attention. 
However, existing methods have limitations in the types of connectivities and delays they are able to handle, for example due to requiring range-parameters for expected delays, or they require computation-intensive procedures, such as computing cross-correlation histograms or performing cross-evaluations of parameter values.  \begin{figure}[t]
  \begin{center}
    \includegraphics[width=\textwidth]{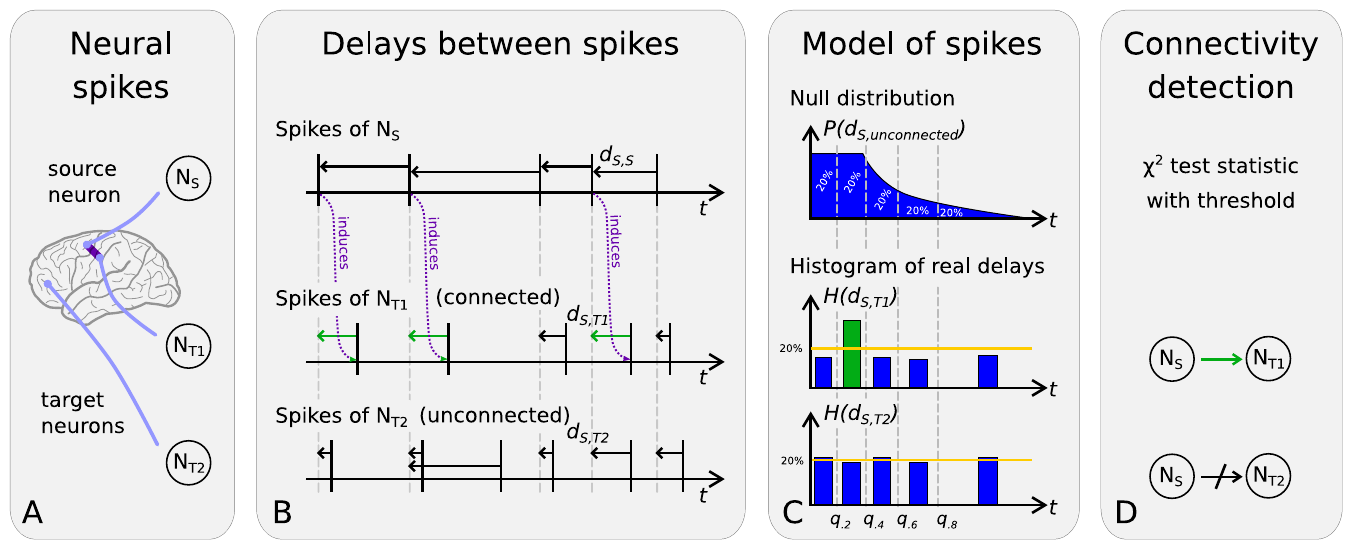} % check!
  \end{center}
  \caption{Schematic visualisation of ACE (\textbf{A}nalysis of \textbf{C}onnectivity of spiking \textbf{E}vents).}
  \label{fig:method}
\end{figure}

We propose a faster and more robust approach for the detection of delayed correlated activity. The principle of this statistical \textbf{A}nalysis of \textbf{C}onnectivity in spiking \textbf{E}vents (ACE) of neurons is illustrated in Fig.~\ref{fig:method}~below. This statistical approach follows the idea of hypothesis testing: Starting (A) with data in the form of spike trains that are recorded for several neurons, the aim is to infer for any pair of source ($N_S$) and target neuron ($N_T$) the pairwise connectivity between them. For this purpose, we compute in step (B) for any pair of neurons the inter-spike intervals, i.e. the delays between their subsequent spikes ($d_{S,S}$, $d_{S,T1}$ and $d_{S,T2}$). 
In the third step (C), we use the inter-spike intervals of a potential source neuron to determine the null distribution of delays for \emph{unconnected} target neurons (shown on top as $P(d_{S,unconnected})$). If a target neuron is \emph{not}connected to the source neuron, the observed distribution of the inter-spike intervals should follow this distribution (shown on the bottom as $H(d_{S,T2})$). In contrast, if this observed distribution (shown at the center as $H(d_{S,T1})$) differs sufficiently, we assume that these two neurons are connected. Thus, in step (D) we use the Pearson's $\chi^2$ test statistic to determine the connectivity for each pair of neurons.

As a consequence, and in contrast to existing methods, our approach works incrementally without using a cross-correlation histogram (faster) and without providing range-parameters for expected delays (more flexible) by assuming that the signal reaches the target neuron faster than the source neuron fires again. This makes it particularly interesting for online processing. 
Summarising, the contributions of our statistical approach for the \textbf{A}nalysis of \textbf{C}onnectivity of spiking \textbf{E}vents (ACE) are:
\begin{itemize}
  \item A statistical, principled approach based on hypothesis testing, by modelling the null distribution of delays between pairs of \emph{unconnected} neurons,
  \item which is fast and robust (no sensitive parameters),
  \item does not assume a particular type of connectivity pattern,
  \item capable to detect connections with unknown, non-discrete delays, that might vary in length between connections.
  \item Based on the distribution of the F1 score for different datasets with similar neuron types (with known ground truth), experiments indicate that the threshold is a robust parameter (transferable to other data sets) (see Fig. \ref{fig:f1-rat}).
\end{itemize}

\section{Related work}\label{sec:rel}
There is a rich literature on analysing neuronal spike train data for inferring connectivity. This comprises recent reviews \cite{MagransdeabrilYoshimotoDoya2018,RoudiDunnHertz2015,Chen2013}, as well as a recent machine learning challenge on neural connectomics \cite{BattagliaEtal2017}. Following \cite{MagransdeabrilYoshimotoDoya2018}, we distinguish model-based approaches from model-free ones. 

An important limitation of model-based approaches is that they rely on assumptions on the data generating process. Thus, their structure and function depend on a large number of factors, resulting in a variety of models and approaches \cite{MagransdeabrilYoshimotoDoya2018}. These include autoregressive models, which are fast but assume a directed linear interaction (e.g., \cite{SmithFallSornborger2011}; generalised linear models, which recently have also been extended to handle transmission delays, although still being limited to small and uniform delays within the network (e.g., \cite{RamirezPaninski2014,ZaytsevMorrisonDeger2015}); and dynamic Bayesian networks, for example \cite{EldawlatlyEtal2010}, where  parameters are learned via simulated annealing, or \cite{PatnaikLaxmanRamakrishnan2011}, which integrates frequent episode mining techniques to improve computational speed. %
This comprises approaches based on the correlation between the activities of neurons, which was the key component in the approach by \cite{SuteraEtal2015}, winning this neural connectomics challenge\cite{BattagliaEtal2017}. While in the simplest form only simultaneous spikes are considered, the extension to cross correlation allows for a delay $\tau$ between spike times \cite{ItoEtal2011}. 
More recently, \cite{SuteraEtal2015} suggested to use the inverse covariance matrix to compute the partial correlation coefficient to measures the conditional dependencies between neurons, thereby filtering out spurious indirect effects.
Extending this idea, \cite{Mohler2014} suggested to use inverse covariance estimates together with an initial convolution filter to preprocess the data. The convolution kernel and other parameter are learned by optimising the binomial log-likelihood function \cite{Mohler2014} on a training data set, where ground truth is known. In the above mentioned challenge they report comparable AUC scores and accuracy compared to the winner, while being noticeably faster. 

A further group of model-free approaches are based on information theoretic approaches. Their applicability for neural connectomics was investigated in \cite{Garofalo2009}, comparing methods based on Mutual Information, Joint-Entropy, Transfer Entropy (TE) and Cross-Correlation. This study revealed Transfer Entropy and Joint-Entropy being the best of the aforementioned methods. Transfer Entropy describes the directed flow of information between two systems \cite{Schreiber2000}. It is equivalent to Granger causality for Gaussian variables \cite{BarnettBarrettSeth2009}, which describes a statistical hypothesis test that measures the ability of predicting future events of a time series based on past events of related time series. However, pairwise Granger causality only detects direct correlations. This leads to problems when two neurons are driven by a common third neuron with different delays \cite{Cadotte2008}. 
Various extensions of Transfer Entropy have been proposed \cite{ItoEtal2011,StetterEtal2012}. 
A Generalised Transfer Entropy approach is proposed in \cite{StetterEtal2012}. For this algorithm, binned spike data is necessary. The underlying assumption is that events in a time bin are correlated to events in earlier time bins. They also consider events in the same time bin and differentiate between a burst and non-burst state of the neural network. Finally, a threshold is used to find correlated neuron pairs.
In \cite{ItoEtal2011}, two different Transfer Entropy-based approaches are proposed, one using Delayed Transfer Entropy (DTE) and one using Higher Order Transfer Entropy (HOTE). The Delayed Transfer Entropy approach calculates the Transfer Entropy for multiple delays (e.g. 1-30ms). HOTE extends the previous approach by considering multiple bins for each delay. 

More recently, statistical methods for analysing synchrony across neurons have been reviewed in \cite{HarrisonAmarasinghamKass2013}. Therein, it is emphasised that the statistical identification of synchronous spiking presumes a null model that describes spiking without synchony. Then, if the observed synchrony is not consistent with the distribution under the null model, i.e. there is more synchony than expected by chance, the null hypothesis of no synchrony is rejected. 
For hypothesis testing, \cite{HarrisonAmarasinghamKass2013} focus their discussion on deriving the test statistics from a cross-correlation histogram (CCH, \cite{PerkelGersteinMoore1967B}). This shows the observed frequency of different time delays between two spike trains, but is scaling dependent and computationally costly.

Finally, the use of Convolutional Neural Networks has been proposed in \cite{Romaszko2015} to learn connectivity directly from calcium imaging data. This approach scored third in the challenge, below the correlation-based approaches, but required extremely high computation time. 
\section{A method for analysing potential connectivity in events}\label{sec:meth} The new method ACE (\textbf{A}nalyser for \textbf{C}orrelated spiking \textbf{E}vents) is a statistical approach following the idea of hypothesis testing. Observing the inter-spike intervals of a potential source neuron, we determine the null distribution of delays for \emph{unconnected} target neurons. If the real observed distribution differ sufficiently (using the Pearson's $\chi^2$ test statistic), we infer that these two neurons are connected. 
In contrast to existing methods, our approach works incrementally without using a cross-correlation histogram (faster) and without providing range-parameters for expected delays (more flexible).

The complete analysis pipeline of our algorithm is provided in Fig.~\ref{fig:method}: In the first step, we compute the delays of two consecutive spikes of all neurons to determine the neuron's parameters $\lambda$ and $RP$ (Sec.~\ref{sec:meth:parameters}, Fig.~\ref{fig:method}B (top)). Thereby, we are able to reconstruct the neuron's delay distribution and to determine the null model for the delay distribution between two neurons~(Sec.~\ref{sec:meth:null}, Fig.~\ref{fig:method}C (top)). As we estimate the null model in advance, we have information on the expected delay distribution and can use the histogram with intervals according to the quantiles to estimate the distribution of observed delays~(Sec.~\ref{sec:meth:quantiles}). Then, the real delays from the source neurons to the target neurons are determined and the histogram is completed (Fig.~\ref{fig:method}C (bottom)). The $\chi^2$ statistic provides a score for distinguishing connected and unconnected neurons (Sec.~\ref{sec:meth:test}, Fig.~\ref{fig:method}D). A threshold is used to determine if the connection score was sufficiently large. A detailed description follows in the next subsections, while the threshold specification is discussed in the evaluation section.
\subsection{Modeling spiking behaviour of a single neuron}\label{sec:meth:parameters}
As we will show in \ref{sec:meth:null}, the null-distribution for the delays between unconnected source and target neurons depends on the distribution of time intervals (delays) $d$ between consecutive spikes of the source neuron~($N_S$). 
To model the random variable ($X_{N_S \rightarrow N_S}$) corresponding to these delays, we use an exponential distribution as a simplification of the gamma distribution, in accordance to \cite{pillow2009time}. This yields two parameters: the refractory period $RP$, describing the time a neuron is inhibited to spike again, and the firing rate $\lambda$, defining the shape of the exponential distribution.
The probability density function~(pdf) of $X_{N_S \rightarrow N_S}= EXP(\lambda) + RP$ and its first two moments are given as:
\begin{align}\label{eq:delays_ns_ns}
  f_{N_S \rightarrow N_S}(d) &= \left\{ { 
    \begin{array}{ll}
      \lambda \exp(-\lambda (d-RP) ) & d \geq RP \\
      0 & d < RP
    \end{array}}\right.\\
  E(X_{N_S \rightarrow N_S}) &= E(EXP(\lambda) + RP) = E(EXP(\lambda)) + RP = \nicefrac{1}{\lambda}+ RP \\
  V(X_{N_S \rightarrow N_S}) &= V(EXP(\lambda) + RP) = V(EXP(\lambda)) = \nicefrac{1}{\lambda^2}
\end{align}

The expectation value $E(X_{N_S \rightarrow N_S})$ and the variance $V(X_{N_S \rightarrow N_S})$ can be incrementally  calculated~\cite{BabcockEtal2003} to find the values for both parameters $RP$ and $\lambda$.
\begin{align}
  &\lambda = \nicefrac{1}{\sqrt{V(X_{N_S \rightarrow N_S})}}& RP = E(X_{N_S \rightarrow N_S}) - \nicefrac{1}{\lambda}
\end{align}\subsection{Determining the null-distribution for uncorrelated neurons}\label{sec:meth:null}
Our approach follows the idea of a statistical test: Instead of deriving models for cases when a source neuron~($N_S$) is connected to a target neuron~($N_T$), we develop a model to describe the delays $d$ if $N_S$ and $N_T$ are \textit{unconnected}. 
Observing a spike at the target neuron at time $t$, and knowing the time $t_S$ of the source neuron's last spike, this delay is $d=t-t_S$.

If the source neuron~($N_S$) is not connected to the target neuron~($N_T$), spikes of $N_T$ seem to appear randomly from the perspective of $N_S$ as they are \textit{independent}. Instead of using the real spike time points, we could also use an equal number of randomly chosen time points. Thus, the null-distribution solely depends on the firing frequency of the source neuron, which is defined by its parameters $RP$ and $\lambda$. 
Determining the distribution of delays $d=t-t_S$ corresponds to estimating the probability $P(X_{N_S \rightarrow N_S}> d)$ that $N_S$ has not spiked again within $[t_s, t]$:
\begin{align}
  P(X_{N_S \rightarrow N_S}> d) = 
  1-&\int_0^{d}f_{N_S \rightarrow N_S}(d') \;\mathrm{d} d' = \left\{ { 
    \begin{array}{ll}
      \exp(-\lambda (d-RP) ) & d \geq RP \\
      1 & 0 \leq d < RP
    \end{array}}\right.\\
  \int_0^{\infty}P(X_{N_S \rightarrow N_S}> d) \;\mathrm{d} d' 
    &= \int_0^{RP}1 d'  + \int_{RP}^{\infty}\exp(-\lambda (d-RP) ) \;\mathrm{d} d' \\
    &= RP + \frac{1}{\lambda}\int_{0}^{\infty}\lambda \exp(-\lambda d ) \;\mathrm{d} d' 
    = RP + \frac{1}{\lambda}
\end{align}
Using the normalised probability from above, we obtain the distribution $X_{N_S \rightarrow N_?}$ of delays between $N_S$ and an unconnected neuron $N_?$:
\begin{align}\label{eq:null-distribution}
f_{N_S \rightarrow N_?}(d) = \frac{P(X_{N_S \rightarrow N_S}> d) }{\int_0^{\infty}P(X_{N_S \rightarrow N_S}> d') \;\mathrm{d} d'}= \left\{ { 
        \begin{array}{ll}
        \frac{\exp(-\lambda (d-RP) )}{RP+ \nicefrac{1}{\lambda}}& d \geq RP \\
        \frac{1}{RP+ \nicefrac{1}{\lambda}}& 0 \leq d < RP
        \end{array}}\right.
\end{align}
Summarising, this null model describes the distribution of delays between two unconnected neurons. Hence, our model is based on (but not similar) to the interspike intervals of neuron $N_S$ which depends on the refractory period ($RP$) and the firing rate ($\lambda$). 

\subsection{Estimating the distribution of observed delays}
\label{sec:meth:quantiles}
To compare the true distribution of observed delays to the distribution under the null model, we build a histogram with $B$ bins such that every bin should contain the same amount of delays following the null distribution. The delay interval of bin $b \in \{1,\dots,B\}$ is given in Eq.~\ref{eq:bin_interval} with $F^{-1}$ being the quantile function (inverse cumulative distribution function) of the null distribution:
\begin{align}\label{eq:bin_interval}
  \mathcal{I}_b = \left[ F^{-1}\left(\frac{b-1}{B}\right),  F^{-1}\left(\frac{b}{B}\right) \right[
\end{align}
Given the source neuron's $RP$ and $\lambda$, this quantile function is (see proof in supplemental material):
\begin{align}\label{eq:quantile_function}
  F^{-1}(q) = \left\{ { 
    \begin{array}{ll}
      RP - \frac{\ln\left(1- (q - \frac{RP}{RP+\nicefrac{1}{\lambda}}) \cdot ( \lambda RP + 1) \right)}{\lambda}& q > \frac{RP}{RP+\nicefrac{1}{\lambda}}\\
      q \cdot (RP+\nicefrac{1}{\lambda}) & q \leq \frac{RP}{RP+\nicefrac{1}{\lambda}}
    \end{array}}\right.
\end{align}

\subsection{Infering connectivity using the Pearson's $\chi^2$-test statistic}
\label{sec:meth:test}

Following the null hypothesis (neurons are not connected), the previously mentioned histogram should be uniformly distributed. Hence, the frequencies $H_b$ of bin $b$ should be similar to $H_b \approx \nicefrac{N}{B}$ ($N = \sum H_b$, which is the total number of delays). 
Our method uses the Pearson's $\chi^2$-Test statistic to find a threshold for distinguishing connected and unconnected neurons.
\begin{align}
  \chi^2 = \sum_{b=1}^{B}\frac{(H_b - \frac{N}{B})^2}{\frac{N}{B}}
\end{align}

Instead of calculating the $p$ value, we directly use the $\chi^2$ statistic to determine a threshold as the degrees of freedom ($B-1$) are similar for every pair of neurons.
\section{Experimental evaluation}\label{sec:eval}

We evaluate our algorithm to the most used baseline techniques~\cite{MagransdeabrilYoshimotoDoya2018} regarding its detection quality and its robustness to parameters like the detection threshold. All code and data are available at our repository\footnote{ \url{https://bitbucket.org/geos/ace-public}}.

\subsection{Baseline algorithms and performance scores}\label{sec:eval:setup}
We compare our algorithm with the method proposed by \cite{Mohler2014}(denoted as \textit{IC}), which is as the winner of the Neural Connectomics Challenge based on inverse covariance but faster, and the higher order transfer entropy (\textit{HOTE}) approach~\cite{ItoEtal2011}. Besides the threshold that all connectivity detection algorithms have in common, both methods require an additional binning parameter that influences the performance. Following the suggestions, we use realistic standard parameters and two variants with 20 and 50 bins. The experiments were implemented in MATLAB (HOTE was provided by the author of \cite{ItoEtal2011}).

As in the challenge and common in literature, we compare the algorithms' connection scores using the Area Under the Precision-Recall Curve (AUPRC)~(see \cite{Powers2011} and \cite{MagransdeabrilYoshimotoDoya2018}). This describes the relationship between precision $ \frac{TP}{TP+FP}$ and recall $ \frac{TP}{TP+FN}$ at different thresholds~\cite{MagransdeabrilYoshimotoDoya2018}.
Additionally we use the F1 score, a standard score for imbalanced data, to equally balance connections and no-connections. Additional similar results for the Area Under the Receiver Operating Characteristic, together with the source code and data sets, are provided in the \textbf{supplemental material}.\subsection{Sensitivity of the algorithms}
To evaluate the sensitivity of algorithms to different neural patterns, in our first series of experiments we used artificially generated data with varying neural parameters, which are summarised in Tab.~\ref{table:performance:AUC}. The ranges of each neural parameter have been chosen according to studies with animals~\cite{izhikevich2006polychronization} and comparable evaluation settings~\cite{MagransdeabrilYoshimotoDoya2018}.

  To evaluate the detection capabilities for each algorithm, we show the \textbf{AUPRC scores} in Tab.~\ref{table:performance:AUC}. Our algorithm outperforms all competitors except for the data set NU\_H with higher number of neurons (equal performances with the HOTE approach) and the data set with high delays (DE\_H) which is more difficult for all algorithms. To explain our poor performance on the latter, we need to recall that the self-initiated firing rate is between $17$ms and $36$ms, calculated as the interval given in expected latency (default $[10, 25)$) plus the RP ($[7, 11)$) for the DE\_H data set. If we observe delays longer than the source neuron's inter-spike intervals (here: $9ms \leq d < 120ms$) it is likely that the source neuron spiked again before its signal reaches the target neuron. Hence, we are not able to link the spike of the source with the spike of the target neuron which makes it impossible to find the respective connection. Fortunately, this behaviour is rare in real neural systems~\cite{izhikevich2006polychronization}(see also Sec.~\ref{sec:eval:real}).

Unfortunately, it is not possible to set the detection threshold of the algorithms to a fixed value. In this section, we aim to evaluate the sensitivity of the respective parameter mentioned in Tab.~\ref{table:performance:AUC}. Therefore, we use each of the three different configurations (low, mid, high) as one fold of an experiment. For each fold, we tune the detection threshold on the remaining folds and calculate the \textbf{F1 score} accordingly. The mean F1 scores and its standard deviation across the folds are presented in Tab.~\ref{tab:performance:F1}. Except for the data set with varying delays, our approach shows superior performance than the baseline algorithms although the AUPRC score differences have not been that large. This indicates that our detection score is more robust to changes in the number of neurons (NU), the latency (LA), the number of connections (CO) and noise (NO). The F1 scores for the delay (DE) data sets for our approach ACE are: DE\_L $0.6164$, DE\_M = ST $0.7539$ and DE\_H = $0.1100$. We see that the previously discussed low performance on DE\_H is the reason for the low mean score.

\begin{figure*}
\begin{minipage}[t]{0.45\textwidth}
		\centering
		\includegraphics[height=4cm]{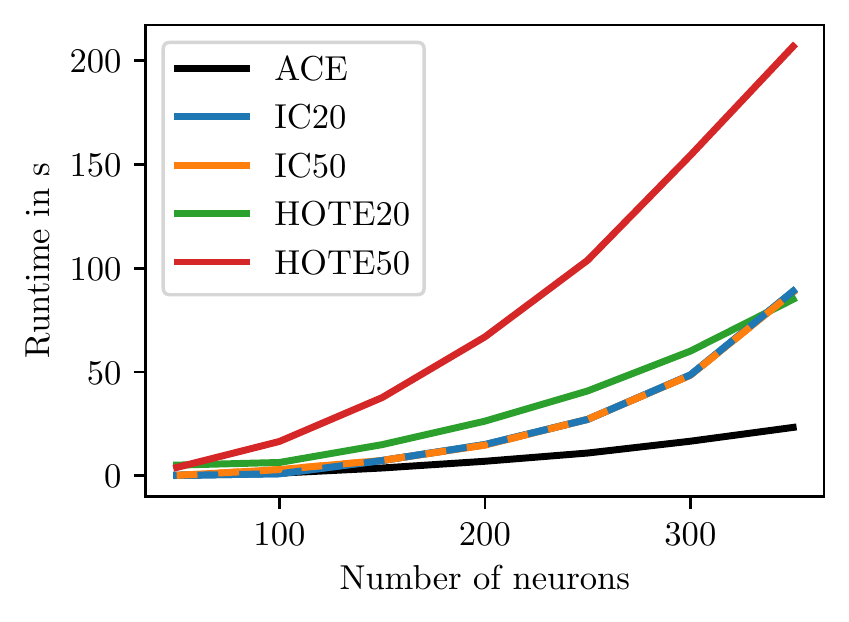} % check!
		\captionof{figure}{Runtime of the strategies over data sets with varying number of neurons.}
		\label{fig:runtime}
\end{minipage}
\begin{minipage}[b]{0.05\textwidth}
	~
\end{minipage}
\begin{minipage}[t]{0.45\textwidth}
		\centering
		\includegraphics[height=4cm]{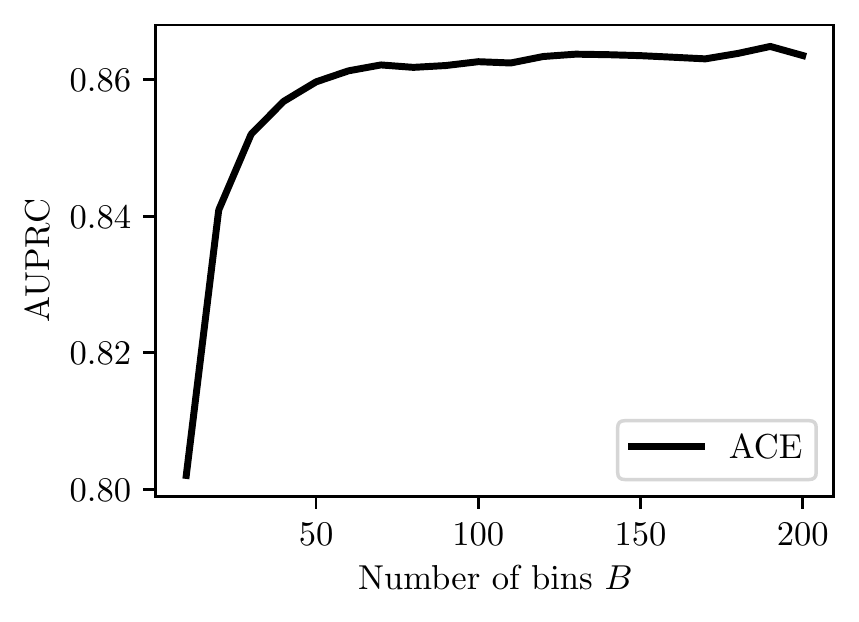} % check!
		\captionof{figure}{AUPRC scores of ACE with varying bin sizes on the ST dataset.}
		\label{fig:parameter-bin}
\end{minipage}
\end{figure*}

In the experiments, we presented two variants of IC and HOTE as they need an additional parameter which highly influences the results. Our \textbf{parameter $B$} has not such an influence on the performance and the runtime (see also Sec.~\ref{sec:eval:complexity}) of ACE. This is visualised in Fig.~\ref{fig:parameter-bin} which plots the AUPRC w.\,r.\,t. the number of bins $B$ on the ST data set. The plot shows that the performance generally increases with higher resolution, but only marginally beyond $B=100$, our default for further experiments.

\begin{table}[h!]
  \centering
  \caption{Area Under the Precision-Recall Curve values for all algorithms on data sets with varying characteristics. The default values for unvaried characteristics are 100 neurons (in the dataset), {[}10, 25) ms + RP expected latency (between two consecutive spikes), 1\% as relative number of connections, {[}5, 9)ms delay (between two connected neurons), and {+0ms}~noise (when determining spike times). The length of the spike stream is $30s$ and the refractory period (RP) is uniformly between $7$ and $11$ ms. The data set name is composed by the characteristic abbreviation and a suffix for either low, mid or high. Thus the setting with 200 Neurons is called NU\_H. }
  \label{table:performance:AUC}
  \centering
  \medskip
  \setlength{\tabcolsep}{3pt}
  \begin{tabular}{llccccc}
    \toprule    
    {}& \textbf{characteristic}& \textbf{ACE}& \textbf{IC20}& \textbf{IC50}& \textbf{HOTE20}& \textbf{HOTE50}\\
    \cmidrule(r){1-7}
    \textbf{ST   }&  defaults & \textbf{0.8626}& 0.2028 & 0.1409 &  0.7832 &  0.7761 \\
    \textbf{NU\_L }& 50 neurons &\textbf{0.8519}& 0.0834 & 0.0616 &  0.6310 &  0.6186 \\
    \textbf{NU\_H }& 200 neurons & 0.9504 & 0.2586 & 0.1038 &  0.9552 &  \textbf{0.9553}\\
    \textbf{LA\_L }& {[}1, 10) ms + RP latency & \textbf{0.8652}& 0.0219 & 0.0138 &  0.5579 &  0.5546 \\
    \textbf{LA\_H }& {[}25, 50) ms + RP latency & \textbf{0.8135}& 0.2500 & 0.1430 &  0.8126 &  0.8119 \\
    \textbf{CO\_L }& 5$\permille$ connections & \textbf{0.8850}& 0.2656 & 0.1620 &  0.8738 &  0.8739 \\
    \textbf{CO\_H }& 2\% connections          & \textbf{0.8086}& 0.0406 & 0.0317 &  0.5952 &  0.5894 \\
    \textbf{DE\_L }& {[}2, 5)ms delay   & 0.7598 & 0.2770 & 0.1528 &  \textbf{0.7671}&  0.7670 \\
    \textbf{DE\_H }& {[}9, 120)ms delay & 0.1334 & 0.1687 & 0.1053 &  0.1505 &  \textbf{0.3669}\\
    \textbf{NO\_M }& +{[}0, 3)ms noise & \textbf{0.7838}& 0.2248 & 0.1225 &  0.7246 &  0.7224 \\
    \textbf{NO\_H }& +{[}0, 5)ms noise & \textbf{0.8518}& 0.2246 & 0.1021 &  0.8322 &  0.8309 \\
    \bottomrule
  \end{tabular}
\end{table}
\begin{table}[h!]
  \caption{Mean F1 score (incl. standard deviations) for all algorithms on the synthetically generated data from Tab.~\ref{table:performance:AUC}. High scores indicate robustness w.\,r.\,t. the varied data set characteristic.}
  \label{tab:performance:F1}
  \centering
  \setlength{\tabcolsep}{4pt}
  \medskip   
    
    \begin{tabular}{lccc}
        \toprule
        {}& \textbf{NU }& \textbf{LA }& \textbf{CO }\\
        \cmidrule{1-4}\\
        \textbf{ACE}& \textbf{0.7663}$\mathit{(\pm 0.08)}$ & \textbf{0.7047}$\mathit{(\pm 0.02)}$ & \textbf{0.7650}$\mathit{(\pm 0.04)}$ \\
        \textbf{IC20}& 0.2430 $\mathit{(\pm 0.11)}$ & 0.1959 $\mathit{(\pm 0.14)}$ & 0.2376 $\mathit{(\pm 0.12)}$ \\
        \textbf{IC50}& 0.1616 $\mathit{(\pm 0.07)}$ & 0.1560 $\mathit{(\pm 0.09)}$ & 0.2013 $\mathit{(\pm 0.09)}$ \\
        \textbf{HOTE20}& 0.6918 $\mathit{(\pm 0.14)}$ & 0.6043 $\mathit{(\pm 0.10)}$ & 0.6882 $\mathit{(\pm 0.12)}$ \\
        \textbf{HOTE50}& 0.6878 $\mathit{(\pm 0.14)}$ & 0.6014 $\mathit{(\pm 0.10)}$ & 0.6851 $\mathit{(\pm 0.12)}$\\
        \bottomrule
    \end{tabular}
    \newline
    \medskip
    \newline
    \begin{tabular}{lcc}
        \toprule
        {}& \textbf{DE }& \textbf{NO }\\
        \cmidrule{1-3}\\
        \textbf{ACE}& 0.4935 $\mathit{(\pm 0.28)}$ & \textbf{0.737}9 $\mathit{(\pm 0.03)}$ \\
        \textbf{IC20}& 0.3419 $\mathit{(\pm 0.04)}$ & 0.3422 $\mathit{(\pm 0.01)}$ \\
        \textbf{IC50}& 0.1853 $\mathit{(\pm 0.05)}$ & 0.1959 $\mathit{(\pm 0.02)}$ \\
        \textbf{HOTE20}& 0.4542 $\mathit{(\pm 0.25)}$ & 0.6406 $\mathit{(\pm 0.02)}$ \\
        \textbf{HOTE50}& \textbf{0.5436}$\mathit{(\pm 0.15)}$ & 0.6376 $\mathit{(\pm 0.02)}$\\
        \bottomrule
    \end{tabular}
\end{table}

\subsection{Computational complexity}
\label{sec:eval:complexity}
To provide a computational run time complexity bound of ACE, let $N$ denote the number of neurons, $M$ the number of spikes over all neurons, and $B$ the number of bins used in the histogram. 
ACE's first step is estimating the refractory period $RP$ and firing rate $\lambda$ for each neuron by iterating once over all its spikes, suming up to $O(M)$ constant time operations for all neurons.
Second, for each neuron's bin the frequencies according to its null model are computed, resulting overall in $O(N \cdot B)$.
Third, the histograms of all neurons are updated after each spike, requiring to insert the observed delay into the corresponding bin. Using a k-d-tree, this requires overall $O(M \cdot N \cdot \log(B))$.
Fourth, the $\chi^2$-test statistic is computed and tested for each pair of neurons, requiring $O(N^2 \cdot B)$ operations.
The last two steps dominate, giving an overall complexity of $O(M \cdot N \cdot \log(B))$ or $O(N^2 \cdot B)$, respectively. In practice, the third step might be the bottleneck, as the number of spikes depends on the number of neurons, i.e., $M > N$, and the number of bins is small (e.g., $B=100$).
In contrast, HOTE has an overall time complexity of $ O(N^2 \cdot (F \cdot D \cdot R + 2^R) ) $, with firing rate $F$, recording duration after discretisation $D$, and $R$ being the total order $(k + l + 1)$ used in the calculations. IC's time complexity is $ O(N^2 \cdot T + N^2 \cdot \log(N) ) $, with $T$ being the number of considered time lags.

Figure \ref{fig:runtime} shows the empirical runtime\footnote{Experiments were performed using a Intel(R) Core(TM) i7-6820HK CPU @ 2.70GHz, 16GB RAM} of all strategies according to the number of neurons in the data set. Those runtimes are obtained by creating datasets that duplicate the spike trains from NE\_L and evaluating them 10 times. One can see that ACE is the fastest. IC20 and IC50 only differ slightly in runtime. HOTE20's runtimes are higher than the ones from IC20 and IC50 for $N \le 300$. HOTE50 is by far the slowest, which shows how HOTE's runtime is affected by the number of bins.

\subsection{Realistically simulated datasets}
\label{sec:eval:real}
In order to have datasets with known ground truth, the standard is to use simulated datasets \cite{MagransdeabrilYoshimotoDoya2018}.
Additionally to the tests on our generated datasets with varying characteristics, we tested the algorithms on an openly available\footnote{See \url{https://crcns.org/data-sets/sim/sim-1/about-sim-1}} spike train dataset with real-world characteristics \cite{BezaireEtal2015}. 
We used the data from the cal\_1x\_exc\_1 experiment split into datasets (RAT) with three folds. To reduce the number of neurons we only considered neurons with more than 100 spikes and at least 5 incoming connections. For each fold, we selected 200 neurons with its corresponding connections. 

\begin{figure}
	\centering
  \includegraphics[width=\textwidth, trim=.7cm 0cm .8cm 0cm, clip]{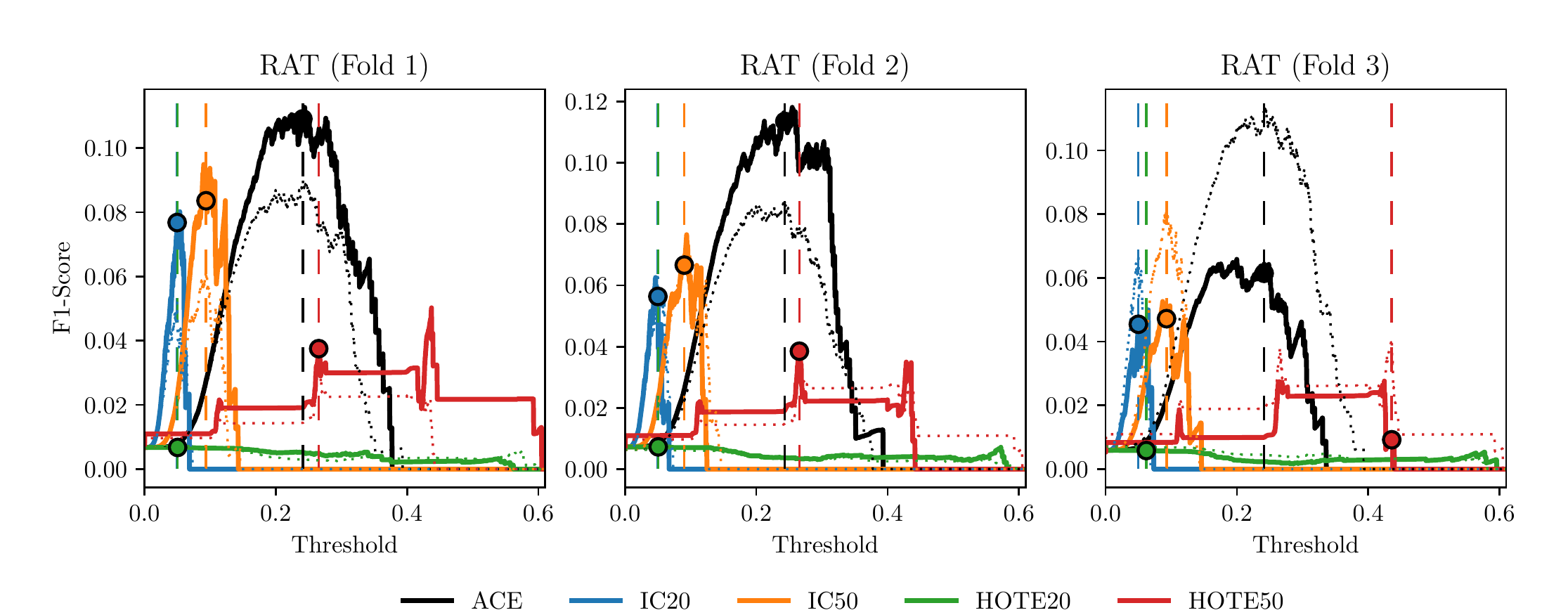} % check!
  \caption{F1 scores with respective to the chosen threshold (here normalised between 0 and 1). The continuous line represents the F1 scores, while the dotted line shows the averaged F1 scores from the other 2 folds. These averaged F1 scores are used to tune the threshold (colored dots, dashed lines).}
  \label{fig:f1-rat}
\end{figure}

The results for these experiments are shown in Tab.~\ref{tab:performance:rat}. ACE outperforms its competitors in terms of AUC and F1 score. The F1 score for each fold w.\,r.\,t. its threshold is shown in Fig.~\ref{fig:f1-rat}. This figure shows that all algorithms with the exception of HOTE 50 show a stable optimal threshold. A wide and high peak  means that non-optimal thresholds are more likely to achieve good performances as can be seen for ACE. A stable shape on the other hand is important as well, as it indicates that similar datasets with known connectivity can be used to tune the threshold.In summary, the experiments align with the theoretical dis-/advantages of our model. The advantages are: 
(1) It does not discretise time. 
(2) By modeling delays of non-connected neurons instead of connected ones, we do not incorporate a bias
from assumptions on delay characteristics. 
(3) The choice of our parameter (number of bins) is not critical, while HOTE can not detect delays exceeding the chosen time window. 
(4) We are faster compared to state-of-the-art algorithms, as no iteration over various delays is required.
\linebreak
In contrast, ACE's theoretical drawbacks are: 
(1) We use of the exponential distribution as a simplification of the gamma distribution with reference to Pillow’s work on the estimation of non-Poisson neural encoding models. 
(2) Each spike of $N_T$ is assigned to the last spike of $N_S$ . If $N_S$ spikes again before the signal reaches $N_T$, the spikes are mismatched and the delay will be miscalculated and the connection might not be found. Fortunately, this is unlikely in real systems \cite{izhikevich2006polychronization}.
(3) Our method (similar to TE and HOTE) does not filter out transitive connections. This requires post-processing which is subject of future work. 

\begin{table}
	\caption{Mean AUROC, AUPRC and F1 score (incl. standard deviations) on the RAT dataset.}
	\label{tab:performance:rat}
	\centering
	\setlength{\tabcolsep}{4pt}
	\medskip   
    \begin{tabular}{lccc}
		\toprule
        {}& \textbf{AUROC }& \textbf{AUPRC }& \textbf{F1 }\\ 
        \cmidrule(r){1-4}
        \textbf{ACE}& \textbf{0.9312 $\mathit{(\pm 0.01)}$}& \textbf{0.0459 $\mathit{(\pm 0.01)}$}& \textbf{0.0947}$\mathit{(\pm 0.02)}$ \\
        \textbf{IC20}& 0.8158 $\mathit{(\pm 0.03)}$          & 0.0194 $\mathit{(\pm 0.00)}$          & 0.0595 $\mathit{(\pm 0.01)}$          \\ 
        \textbf{IC50}& 0.8864 $\mathit{(\pm 0.02)}$          & 0.0284 $\mathit{(\pm 0.01)}$          & 0.0658 $\mathit{(\pm 0.01)}$          \\ 
        \textbf{HOTE20}& 0.3042 $\mathit{(\pm 0.05)}$          & 0.0022 $\mathit{(\pm 0.00)}$          & 0.0066 $\mathit{(\pm 0.00)}$          \\ 
        \textbf{HOTE50}& 0.6568 $\mathit{(\pm 0.03)}$          & 0.0090 $\mathit{(\pm 0.00)}$          & 0.0284 $\mathit{(\pm 0.01)}$          \\ 
		\bottomrule
\end{tabular}
\end{table}

\section{Conclusion}\label{sec:conc}
This paper addressed the problem of detecting delayed connectivity of neurons. An approach named ACE for detecting correlated, but delayed spike event patterns in streams was introduced. Following the principle of hypothesis testing, the approach is based on a null model for the distributon of inter-spike delays for \emph{uncorrelated}~neurons. Against this null-distribution, the distribution of observed inter-spike delays is compared using a Pearson's $\chi^2$ test statistic.

In an experimental evaluation, this algorithm was compared against recently proposed approaches based on inverse covariance and higher order transfer entropy, on two types of data sets: For realistic benchmarking, data sets based on the publicly available, state-of-the art CRCNS data generator, and for analysing the robustness, data sets with varying characteristics based on our own data generator. 
On all data sets, ACE is faster and performs comparable or better in terms of AUPRC, F1 and AUROC (see supplemental material) score, except for networks with very long inter-spike delays that interfere with uncorrelated spike activity. In particular, ACE performs also better on the publicly available, state-of-the-art benchmark data generator with realistic spike characteristics. ACE has only two parameters, both being very robust and transferable between data sets of similar characteristic. ACE is fast and flexible, allowing to detect connections with a priori unknown, non-discrete delays, that might vary in length between connections. Furthermore, due to its incremental nature, ACE has potential for being used in online processing.

\nocite{MuinoEtal2013}
\bibliographystyle{alpha}
\bibliography{ace_long}

\section*{Appendix}
\subsection{Proof of Equation \ref{eq:quantile_function}}
\label{app:proof_quantile_funct}
In this section, we show the derivation of the quantile function $F^{-1}(q)$ of the null-distribution $f(x)$:

\begin{align}
f_{N_S \rightarrow N_?}(d)
&= \left\{ { 
\begin{array}{ll}
\frac{\exp(-\lambda (d-RP) )}{RP+ \nicefrac{1}{\lambda}}& d \geq RP \\
\frac{1}{RP+ \nicefrac{1}{\lambda}}& 0 \leq d < RP
\end{array}}\right.
\end{align}

\pagebreak[4]
The cumulative probability function $F(x)$ is given as follows:
\begin{enumerate}
 \item  $\forall x \leq RP$:
    \begin{align}
    F(x) = \int_0^x \frac{1}{RP+ \nicefrac{1}{\lambda}}\mathrm{d}t =
    \frac{x}{RP+ \nicefrac{1}{\lambda}}
    \end{align}

  \item  $\forall x > RP$:
    \begin{align}
    F(x) =& \frac{RP}{RP + \nicefrac{1}{\lambda}}+
    \int_{RP}^x \frac{\exp(-\lambda (t-RP))}{RP+ \nicefrac{1}{\lambda}}\mathrm{d}t\\
    =& \frac{RP}{RP + \nicefrac{1}{\lambda}}+ 
    \frac{1}{RP + \nicefrac{1}{\lambda}}
    \int_{0}^{x-RP}\exp(-\lambda t)\\
    =& \frac{RP}{RP + \nicefrac{1}{\lambda}}+ 
    \frac{1}{RP + \nicefrac{1}{\lambda}}\cdot
    \left[-\frac{\exp(-\lambda t)}{\lambda}\right]_{0}^{x-RP}\\
    =& \frac{RP}{RP + \nicefrac{1}{\lambda}}+ 
    \frac{1}{RP + \nicefrac{1}{\lambda}}\cdot 
    \left[ - \frac{\exp(-\lambda(x-RP))}{\lambda}+ \frac{1}{x}\right]\\
    =& \frac{1}{RP + \nicefrac{1}{\lambda}}\cdot
    \left[RP + \frac{1 - \exp(-\lambda (x - RP))}{\lambda}\right]
    \end{align}
\end{enumerate}

The quantile function $F^{-1}(q)$ is given as:
\begin{enumerate}
 \item  $\forall q \leq \frac{RP}{RP + \nicefrac{1}{\lambda}}$:
    \begin{align}
    q &= \frac{x}{RP+ \nicefrac{1}{\lambda}}\\
    x &= q (RP+ \nicefrac{1}{\lambda})
    \end{align}

 \item  $\forall q > \frac{RP}{RP + \nicefrac{1}{\lambda}}$:
    \begin{align}
    q &= \frac{RP + \nicefrac{1}{\lambda}- \frac{\exp(-\lambda(x-RP))}{\lambda}}{RP + \nicefrac{1}{\lambda}}\\
    q \cdot (RP + \nicefrac{1}{\lambda}) - RP - \nicefrac{1}{\lambda}
    &= - \frac{\exp(-\lambda(x-RP))}{\lambda}\\
    \exp\left( -\lambda \left( x-RP\right) \right) 
    &=-\lambda \left( q\left( RP+\nicefrac {1}{\lambda }\right) -RP-\nicefrac {1}{\lambda }\right) \\
    -\lambda(x-RP) &= \ln \left( -\lambda \left( q ( RP+ \nicefrac{1}{\lambda}) - RP - \nicefrac{1}{\lambda}\right)\right) \\
    x &=\frac {\ln \left( -\lambda \left( q\left( RP+ \ \nicefrac{1}{\lambda}
    \right) -\left( RP- \nicefrac{1}{\lambda}\right) \right) \right) }{-\lambda }+RP \\
    & = RP-\frac {\ln \left( \left ( \lambda \cdot RP+1\right) \left( 1+q\right) \right) }{\lambda }
    \end{align}
\end{enumerate}\end{document}